\documentclass{jpp}
\usepackage{graphicx}
\usepackage{epstopdf, epsfig}
\usepackage{amsmath}
\usepackage{color} 

\newcommand{\ex}{\mathbf{e}_{\rm x}}
\newcommand{\ey}{\mathbf{e}_{\rm y}}
\newcommand{\ez}{\mathbf{e}_{\rm z}}

\shorttitle{Relativistic particle mover}
\shortauthor{J. P\'etri}

\title[Fully implicit relativistic particle pusher]{A fully implicit numerical integration of the relativistic particle equation of motion}

\author{J. P\'etri\aff{1}
  \corresp{\email{jerome.petri@astro.unistra.fr}}}

\affiliation{\aff{1}Universit\'e de Strasbourg, CNRS, Observatoire astronomique de Strasbourg, UMR 7550, F-67000 Strasbourg, France.}

\begin{document}

\maketitle

\begin{abstract}
Relativistic strongly magnetized plasmas are produced in laboratories thanks to state-of-the-art laser technology but can naturally be found around compact objects such as neutron stars and black holes. Detailed studies of the behaviour of relativistic plasmas require accurate computations able to catch the full spatial and temporal dynamics of the system. Numerical simulations of ultra-relativistic plasmas face severe restrictions due to limitations in the maximum possible Lorentz factors that current algorithm can reproduce to good accuracy. In order to circumvent this flaw and repel the limit to $\gamma\approx10^9$, we design a new fully implicit scheme to solve the relativistic particle equation of motion in an external electromagnetic field using a three dimensional Cartesian geometry. We show some examples of numerical integrations in constant electromagnetic fields to prove the efficiency of our algorithm. The code is also able to follow the electric drift motion for high Lorentz factors. In the most general case of spatially and temporally varying electromagnetic fields, the code performs extremely well as shown by comparison with exact analytical solutions for the relativistic electrostatic Kepler problem as well as for linearly and circularly polarized plane waves.
\end{abstract}

\section{Introduction}

Relativistic magnetized flows arise in many astrophysical contexts, especially around compact objects such as neutron stars and black holes. Plasma regimes are typically collisionless and comprise a substantial fraction of electron-positron pairs. Numerical techniques to investigate the behaviour of such plasmas require a kinetic description. Unfortunately solving the full Vlasov equations in 6D (3 space and 3 velocity directions) on a computer remains out of reach with current technology. Therefore Particle in Cell (PIC) methods are preferred and well suited for numerically solving collisionless plasma physics problems where kinetic effects prevail. Nevertheless, particle methods are inherently prone to numerical noise. Moreover for ultra strong magnetic fields such as those anchored into pulsars and magnetars, the time scale of gyration is much too small compared to the long term evolution of the plasma rendering computations on dynamical time scales impossible.

The magnetospheres of compact objects, black holes and neutron stars, are filled with relativistic electron-positron pairs travelling in a ultra strong magnetic field close to or even above the critical magnetic field of $4.4\times10^9$~T where quantum electrodynamics processes become dominant. Moreover the plasma evolves in an almost perfectly collisionless regime where coulombian collisions can be neglected. These plasmas are prone to several instabilities and reconnection phenomena as well as radiative processes that need to be studied within a kinetic description of the fluid. 

The Vlasov approach represents the best mean to accurately investigate the long term photon-plasma bath evolution. However it requires integration in a six dimensional phase space, computationally very expensive and unfortunately impossible to treat on contemporary computers. Another less demanding alternative is given by Particle In Cell (PIC) simulations that sample the phase space into millions or billions of individual particles evolving in time via macroscopic electromagnetic interactions (no particle-particle interaction taken into account in the simplest approach). Both techniques require a numerical integration of the particle equation of motion in an external electromagnetic field. Performance, advantages and drawbacks of PIC methods are extensively exposed in the standard textbooks on particle simulations of plasmas such as \cite{birdsall_plasma_2005} and \cite{hockney_computer_1988}. In any case, it is usually difficult to solve this problem in the ultra-relativistic regime where particles can reach Lorentz factor larger than several millions or billions $\gamma>10^9$. At these energies, radiation reaction is expected to play a prominent role on the long term dynamics. In the non relativistic regime the Boris method \citep{boris_relativistic_1970} and the leapfrog method are classical explicit schemes often quoted. Boris algorithm has been extended to relativistic motion by \cite{vay_simulation_2008}. A volume-preserving relativistic integrator was recently developed by \cite{higuera_structure-preserving_2017}. Splitting techniques for high order symmetric volume-preserving methods can be found in \cite{he_yang_high_2016}. A relativistic version of the leapfrog method was also implemented in the code of \cite{melzani_apar-t:_2013} but with possible issues for ultra-relativistic speeds. Realistic simulations have to take care of exact particle energy conservation which is a concern of all available methods \citep{lapenta_particle_2011}. Flaws already encountered in non relativistic plasma simulations are enhanced in relativistic regimes. \cite{vay_modeling_2014} recently reviewed about these relativistic PIC methods. 

Simulations of plasmas around neutron stars are particularly demanding because of the extreme conditions in the vicinity of such stars. Indeed a typical ratio between the cyclotron frequency~$\omega_{\rm B}$ and the stellar rotation frequency~$\Omega$ is
\begin{equation}
 \frac{\omega_{\rm B}}{\Omega} = \frac{e\,B}{m\,\Omega} \approx 10^{16-19}.
\end{equation}
Moreover the ratio between neutron star radius~$R$ and Larmor radius~$r_{\rm B}$ is also large
\begin{equation}
 \frac{R}{r_{\rm B}} = \frac{R\,\omega_{\rm B}}{c} = \frac{\omega_{\rm B}}{\Omega} \, \frac{R}{r_{\rm L}} \approx \frac{10^{15}}{\gamma} \, \frac{R}{r_{\rm L}}
\end{equation}
where $r_{\rm L}=c/\Omega \approx 10-10.000\,R$ is the light-cylinder radius, i.e. another important length scale indicating the transition between the static quasi-stationary and the wave zone induced by rotating the magnet anchored into the neutron star. $\gamma$ is the particle Lorentz factor. Leptons extracted from the neutron star surface are expected to start with modest kinetic energy, $\gamma\gtrsim1$, thus the most severe constrain on this ratio is obtained for $\gamma=1$. Such huge gaps need to be taken into account for realistic modelling of pulsars. In this communication, we show how to efficiently and accurately solve for this motion for any Lorentz factor up to at least $\gamma=10^9$ in some cases. We first remind how to implicitly solve the equation of motion in the non relativistic and relativistic case and how update implicitly the particle position. Then we test our algorithm in some special configurations with constant external electromagnetic fields and with spatially and temporally varying fields.

\section{Equation of motion}

The equation of motion for a relativistic charged particle of mass $m$ and charge $q$, evolving in an external electromagnetic field with $\mathbf{E}$ the electric part and $ \mathbf{B}$ the magnetic part, is given by the Lorentz force according to
\begin{subequations}
\begin{align}
  \label{eq:Vitesse}
  \frac{d\mathbf{r}}{dt} & = \mathbf{v} \\
  \label{eq:Force_Lorentz}
  \frac{d\mathbf{p}}{dt} & = q \, ( \mathbf{E} + \mathbf{v} \wedge \mathbf{B}) .
\end{align}
\end{subequations}
The momentum is deduced from the three velocity $\mathbf{v}$ or from the four-velocity $\mathbf{u}$ by
\begin{equation}
  \mathbf{p} = \gamma \, m \, \mathbf{v} = m \, \mathbf u .
\end{equation}
The particle position is denoted by the vector $\mathbf{r}$ and $t$ is the time as measured by an observer attached to a frame at rest. In order to show the complication arising in the relativistic regime, we first remind the non relativistic implicit algorithm to solve for the particle velocity. Then we extend to the relativistic case and show how to deal with the Lorentz factor.

\subsection{Non relativistic motion}

In the non-relativistic limit, the Lorentz factor is $\gamma=1$ and the momentum simplifies into $\mathbf{p}=m\,\mathbf{v}$. The equation of motion therefore becomes 
\begin{equation}
  \label{eq:Force_Lorentz_Non_Rel}
  \frac{d\mathbf{v}}{dt} = \frac{q}{m} \, ( \mathbf{E} + \mathbf{v} \wedge \mathbf{B}) .
\end{equation}
We look for an implicit method to solve this equation. The discretized version of the Lorentz force Eq.~(\ref{eq:Force_Lorentz_Non_Rel}) is
\begin{equation}
  \frac{\mathbf{v}^{n+1/2} - \mathbf{v}^{n-1/2}}{\Delta t} = \frac{q}{m} \, \left( \mathbf{E}^{\,n} + \frac{\mathbf{v}^{n+1/2} + \mathbf{v}^{n-1/2}}{2} \wedge \mathbf{B}^{\,n} \right) .
\end{equation}
We rearrange terms in order to collect expressions containing only the unknown velocity and the next time step $\mathbf{v}^{n+1/2}$ by writing
\begin{equation}
  \mathbf{v}^{n+1/2} - \mathbf{v}^{n+1/2} \wedge \frac{q\,\Delta t\,\mathbf{B}^{\,n}}{2\,m} =  \mathbf{v}^{n-1/2} + \mathbf{v}^{n-1/2} \wedge \frac{q\,\Delta t\,\mathbf{B}^{\,n}}{2\,m} + \frac{q\,\Delta t \, \mathbf{E}^{\,n}}{m} .
\end{equation}
This system can be solved explicitly in terms of $\mathbf{v}^{n+1/2}$ by introducing two constant parameters such that
\begin{eqnarray}
  \label{eq:alpha}
  \alpha = \frac{q \, \Delta t}{m} & ; & p = \frac{\alpha}{2} .
\end{eqnarray}
We also define the matrix for the magnetic rotation
\begin{equation}
  \label{eq:MatriceRn}
  R^n =
    \begin{bmatrix}
      0          &   p \, B_z^n & - p \, B_y^n \\
      - p \, B_z^n &   0        &   p \, B_x^n \\
      p \, B_y^n & - p \, B_x^n & 0
    \end{bmatrix} 
\end{equation}
and the matrix for the electric acceleration
\begin{equation}
  \label{eq:MatriceSn}
  S^n = 
    \begin{bmatrix}
      \alpha \, E_x^n \\
      \alpha \, E_y^n \\
      \alpha \, E_z^n
    \end{bmatrix} .
\end{equation}
In matrix notation, the discretized equation of motion becomes
\begin{equation}
  \label{eq:Resolution_Vn12}
  ( I_3 - R^n ) \, V^{n+1/2} = ( I _3 + R^n ) \, V^{n-1/2} + S^n  .
\end{equation}
This system is solved for the unknown vector $V^{n+1/2}$ to deduce the velocity at the next time $t^{n+1/2}$ by
\begin{equation}
  V^{n+1/2} = ( I_3 - R^n )^{-1} \, \left[ ( I _3 + R^n ) \, V^{n-1/2} + S^n \right] .
\end{equation}
The inverse matrix of $( I_3 - R^n )$ is easily found to be 
\begin{equation}
  \label{eq:MatriceRnInverse}
  ( I_3 - R^n )^{-1} = \frac{1}{1 + p^2 \, B^2} \, 
    \begin{bmatrix}
      1 + B_x^2 \, p^2 & B_x \, B_y \, p^2 + B_z \, p & B_x \, B_z \, p^2 - B_y \, p \\
      B_x \, B_y \, p^2 - B_z \, p & 1 + B_y^2 \, p^2 & B_y \, B_z \, p^2 + B_x \, p \\
      B_x \, B_z \, p^2 + B_y \, p & B_y \, B_z \, p^2 - B_x \, p & 1 + B_z^2 \, p^2
    \end{bmatrix} .
\end{equation}
$V^{n+1/2}$ can be computed for any time step at the price of a costly matrix multiplication. The inversion of the matrix in front of $V^{n+1/2}$ is the crucial step to an implicit integration of the equation of motion. In the relativistic case, the procedure is far from trivial because the Lorentz factor is a priori not known at the next time step. However, as we show in the next section, it is possible to find the Lorentz factor by solving a biquadratic equation.

\subsection{Relativistic motion}

The relativistic equation of motion is treated in a similar manner as before. However, a complication stems from the indeterminacy of the particle Lorentz factor at the next time $t^{n+1/2}$. Indeed, the discretized version of the Lorentz force Eq.~(\ref{eq:Force_Lorentz}) is
\begin{equation}
  \frac{\gamma^{n+1/2} \, \mathbf{v}^{n+1/2} - \gamma^{n-1/2} \, \mathbf{v}^{n-1/2}}{\Delta t} = \frac{q}{m} \, \left( \mathbf{E}^{\,n} + \frac{\mathbf{v}^{n+1/2} + \mathbf{v}^{n-1/2}}{2} \wedge \mathbf{B}^{\,n} \right) .
\end{equation}
Written in terms of the four-velocity we get
\begin{equation}
  \frac{\mathbf{u}^{n+1/2} - \mathbf{u}^{n-1/2}}{\Delta t} = \frac{q}{m} \, \left[ \mathbf{E}^{\,n} + \frac{1}{2} \, \left( \frac{\mathbf{u}^{n+1/2}}{\gamma^{n+1/2}} + \frac{\mathbf{u}^{n-1/2}}{\gamma^{n-1/2}} \right) \wedge \mathbf{B}^{\,n} \right] .
\end{equation}
This version has the advantage of being fully symmetric in time, therefore conserves energy and is moreover Lorentz invariant. This is the procedure chosen by \cite{vay_simulation_2008}. The unknown velocity at time $t^{n+1/2}$ is separated from the known velocity at time $t^{n-1/2}$ such that $\mathbf{u}^{n+1/2}$ has to be found according to the relation
\begin{equation}
  \label{eq:Force_Lorentz_Rel_Discret}
  \mathbf{u}^{n+1/2} - \frac{\mathbf{u}^{n+1/2}}{\gamma^{n+1/2}} \wedge \frac{q\,\Delta t\,\mathbf{B}^{\,n}}{2\,m} =  \mathbf{u}^{n-1/2} + \frac{\mathbf{u}^{n-1/2}}{\gamma^{n-1/2}} \wedge \frac{q\,\Delta t\,\mathbf{B}^{\,n}}{2\,m} + \frac{q\,\Delta t \, \mathbf{E}^{\,n}}{m} .
\end{equation}
Contrary to the non relativistic case, we cannot solve explicitly for $\mathbf{u}^{n+1/2}$ because of the Lorentz factor $\gamma^{n+1/2}$ that remains unknown. However it is linked to the sought velocity $\mathbf{u}^{n+1/2}$ by 
\begin{equation}
\label{eq:gammafnu}
(\gamma^{n+1/2})^2 = 1 + (u^{n+1/2}/c)^2 .
\end{equation} 
Fortunately, this factor can be determined explicitly analytically following the method outlined in the next lines. Let us introduce the same variables as before defined by eq.~(\ref{eq:alpha}) and two new parameters taking into account the actual Lorentz factor
\begin{equation}
 \label{eq:pmoins_pplus}
  p^\pm = \frac{p}{\gamma^{n\pm1/2}} 
\end{equation}
as well as the matrices $R^n$ and $S^n$. Compared to the non-relativistic situation, we need to replace the matrix $R^n$ by introducing the Lorentz factor to get the useful matrices as $R^n/\gamma^{n+1/2}$ for $U^{n+1/2}$ and as $R^n/\gamma^{n-1/2}$ for $U^{n-1/2}$. The four-velocity at time $t^{n+1/2}$ is solution of the matrix system
\begin{equation}
  \left(  I_3 - \frac{R^n}{\gamma^{n+1/2}} \right)  \, U^{n+1/2} = \left( I_3 + \frac{R^n}{\gamma^{n-1/2}} \right) \, U^{n-1/2} + S^n .
\end{equation}
We invert the matrix $\left( I_3 - \frac{R^n}{\gamma^{n+1/2}}\right) $ to obtain
\begin{eqnarray}
  \left(  I_3 - \frac{R^n}{\gamma^{n+1/2}}\right)^{-1} & = & 
  \frac{1}{1 + (p^+)^2 \, B^2} \, \\
  & \times & \left[ 
    \begin{array}{lll}
      B_x^2 \, (p^+)^2 + 1 & B_x \, B_y \, (p^+)^2 + B_z \, p^+ & B_x \, B_z \, (p^+)^2 - B_y \, p^+ \\
      B_x \, B_y \, (p^+)^2 - B_z \, p^+ & B_y^2 \, (p^+)^2 + 1 & B_y \, B_z \, (p^+)^2 + B_x \, p^+ \\
      B_x \, B_z \, (p^+)^2 + B_y \, p^+ & B_y \, B_z \, (p^+)^2 - B_x \, p^+ & B_z^2 \, (p^+)^2 + 1
    \end{array}
  \right] \nonumber \\
  {\rm det} \left(I_3 - \frac{R^n}{\gamma^{n+1/2}} \right) & = & 1 + (p^+)^2 \, B^2  .
\end{eqnarray}
The matrix can not be singular because its determinant is always larger than unity. The four velocity is derived at the next time step $t^{n+1/2}$ according to
\begin{equation}
  \label{eq:Resolution_Un12}
  U^{n+1/2} = ( I_3 - \frac{R^n}{\gamma^{n+1/2}} )^{-1} \, \left[ ( I_3 + \frac{R^n}{\gamma^{n-1/2}} ) \, U^{n-1/2} + S^n \right] .
\end{equation}
The Lorentz factor $\gamma^{n+1/2}$ remains unknown but it can be deduced from the solution of a biquadratic equation. Indeed, from the relation between four-velocity and Lorentz factor as given in eq.~(\ref{eq:gammafnu}), eq.~(\ref{eq:Resolution_Un12}) allows one to write the Lorentz factor as a root of the biquadratic polynomial defined by
\begin{equation}
  \label{eq:Quadratique_Gamma}
  (\gamma^{n+1/2})^4 + a_1 \, (\gamma^{n+1/2})^2 + a_2 = 0 .
\end{equation}
The coefficients are given by
\begin{subequations}
  \label{eq:Coeff_Quadr}
\begin{align}
  a_2 & =-p^2 \, \left[ ({B^n})^2 + \left\lbrace ( \mathbf{S}^n + \mathbf{u}^{n-1/2} ) \cdot \mathbf{B}^n \right\rbrace ^2 \right] \\
  a_1 & = p^2 \, ({B^n})^2 - 1 - (\mathbf{S}^n + \mathbf{u}^{n-1/2})^2 - p^2 \, (\mathbf{v}^{n-1/2} \wedge \mathbf{B}^n)^2 + 2 \, p \, ( \mathbf{v}^{n-1/2} \wedge \mathbf{S}^n ) \cdot \mathbf{B}^n \nonumber \\
  & = p^2 \, ({B^n})^2 - 1 - (\mathbf{u}^{n-1/2})^2 - 2 \, \mathbf{S}^n \cdot \mathbf{u}^{n-1/2} - ( p \, \mathbf{v}^{n-1/2} \wedge \mathbf{B}^n + \mathbf{S}^n )^2 .
\end{align}
\end{subequations}
The are formally four solutions for the Lorentz factor, two of them being negative and rejected, and only one satisfying the physical condition $\gamma^{n+1/2}\geqslant1$. It is explicitly given by
\begin{equation}
\gamma^{n+1/2} = \sqrt{\frac{\sqrt{a_1^2-4\,a_2}-a_1}{2}} .
\end{equation}
This Lorentz factor is inserted into the matrix $( I_3 - \frac{R^n}{\gamma^{n+1/2}} )^{-1}$ to deduce the four-velocity at the next time step $t^{n+1/2}$.

This analytical procedure to solve for the implicit discretization of the momentum equation has already been given by \cite{vay_simulation_2008} although in a different form. In order to retrieve his results, we conform to his notation and introduce the following quantities
\begin{subequations}
\begin{align}
 \mathbf{\tau} & = p \, \mathbf{B}^n \\
 \mathbf{t} & = \frac{\mathbf{\tau}}{\gamma^{n+1/2}} = p^+ \, \mathbf{B}^n \\
 \mathbf{u}' & = \mathbf{u}^{n-1/2} + \frac{\mathbf{u}^{n-1/2}}{\gamma^{n-1/2}} \wedge \mathbf{\tau} + \mathbf{S}^n .
\end{align}
\end{subequations}
The updated velocity is therefore
\begin{equation}
\mathbf{u}^{n+1/2} = \mathbf{u}' + \mathbf{u}^{n+1/2} \wedge \mathbf{t}  .
\end{equation}
Taking the dot and cross product with $\mathbf{t}$ we get
\begin{equation}
 \mathbf{u}^{n+1/2} = \frac{1}{1+t^2} \, ( \mathbf{u}' + \mathbf{u}' \wedge \mathbf{t} + (\mathbf{u}' \cdot \mathbf{t}) \, \mathbf{t} )
\end{equation}
as found by \cite{vay_simulation_2008}. In matrix form this reduces to
\begin{equation}
\mathbf{u}^{n+1/2} = \frac{1}{1+t^2} \, 
\begin{bmatrix}
      1+t_x^2 & t_x \, t_y + t_z & t_x \, t_z - t_y \\
      t_x \, t_y - t_z & 1 + t_y^2 & t_y \, t_z + t_x \\
      t_x \, t_z + t_y & t_y \, t_z - t_x & 1 + t_z^2
    \end{bmatrix}
\end{equation}
which is the matrix form eq.~(\ref{eq:Resolution_Un12}). Adapting our notation to expressions given by \cite{vay_simulation_2008}, our biquadratic equation for the new Lorentz factor reduces exactly to the same expression. Consequently, the analytic expressions we found, although formulated in a matrix form, are at the end exactly the same as in \cite{vay_simulation_2008} who uses a vector form instead.

\subsection{Position update and fully implicit scheme}

\cite{vay_simulation_2008} algorithm treats the velocity part implicitly but evolves the position explicitly in a leapfrog fashion. A more judicious scheme would treat both position and velocity updates implicitly, with a similar discretization. Such modification gets rid of the staggered grids in time where positions are evaluated at full time steps whereas velocities are computed at half time steps or vice-versa. Fully implicit schemes are usually computationally expensive but in the present work we are mostly concerned with precision, accuracy and stability of the integration technique. This is particularly relevant for particles moving in ultra strong electromagnetic fields like those met in pulsar magnetospheres.

The new scheme we propose is able to handle spatially and temporally varying fields to high accuracy. The position and velocity updates are given similarly by
\begin{subequations}
\label{eq:Picard}
\begin{align}
 \frac{\mathbf{r}^{n+1} - \mathbf{r}^n}{\Delta t} & = \mathbf{v}^* \\
 \frac{\mathbf{u}^{n+1} - \mathbf{u}^n}{\Delta t} & = \frac{q}{m} \, [ \mathbf{E}(\mathbf{r}^*,t^*) + \mathbf{v}^* \wedge \mathbf{B}(\mathbf{r}^*,t^*) ] \\ 
 \mathbf{r}^* & = \frac{\mathbf{r}^{n+1} + \mathbf{r}^n}{2} \\
 \mathbf{v}^* & = \frac{\mathbf{v}^{n+1} + \mathbf{v}^n}{2} = \frac{1}{2} \, \left( \frac{\mathbf{u}^{n+1}}{\gamma^{n+1}} + \frac{\mathbf{u}^{n}}{\gamma^{n}}\right) \\
 t^* & = \frac{t^{n+1} + t^n}{2} .
\end{align}
\end{subequations}
The derivatives on the right hand side for position and velocity updates are evaluated at the mean time~$t^*$ between $t^n$ and $t^{n+1}$. The external electromagnetic field is evaluated at the mean position~$\mathbf{r}^*$ between $\mathbf{r}^n$ and $\mathbf{r}^{n+1}$.

A direct solution of this non-linear system is usually too costly. Another technique to solve it uses the Picard iteration method. Starting from a first guess of the solution given by $\mathbf{r}^{n+1} = \mathbf{r}^{n}$ and $\mathbf{u}^{n+1} = \mathbf{u}^{n}$, the position and velocity are updated at time $t^{n+1}$ according to eq.~(\ref{eq:Picard}) until convergence to a prescribed accuracy. When the difference between two successive updates becomes less than a user prescribed tolerance, convergence is achieved. The number of iterations required is usually less than ten.

In the next section we provide some numerical examples of particle trajectories in constant and spatially/temporally varying electromagnetic fields to check the efficiency and accuracy of our algorithm.

\section{Special cases}

The solution of the system simplifies drastically in special cases where only a magnetic or an electric field is present. For the purpose of numerical tests, we use normalized units and follow the motion of an electron such that its mass is $m=1$ and its charge $q=-1$ in these units. The strength of the electric or magnetic field is also normalized to values taken to be in the set $\{10^{-3},1,10^3\}$. The matrices in the algorithm contain products between different Cartesian components of the electromagnetic field. In order to test extensively the code, we chose to integrate numerically along each coordinate axis $\{x,y,z\}$ separately but also along a combination of two or even the three coordinate axes in order to get the most general matrices with all elements computed to give non trivial values. Comparing the simulation outputs with exact analytical expressions is easy in a constant and homogeneous electromagnetic field. Interestingly, some exact expressions are also known for a monopolar electric field and for polarized plane waves. Comparison between numerical and analytical solutions are given below for special electromagnetic field configurations.

\subsection{Constant electric field}

Let us assume a constant electric field along the $x$ axis $\mathbf{E} = E\,\ex$ and a particle initially at position $x=0$ at time $t=0$ with zero initial velocity. The particle is subject to a constant acceleration (in his rest frame) in the $x$ direction with intensity $g=qE/m$. Moreover its trajectory can be exactly integrated giving
\begin{subequations}
\begin{align}
 x & = \frac{c^2}{g} \, \left[ \sqrt{1+\left( \frac{g\,t}{c} \right)^2} - 1 \right] \\
 v & = \frac{g\,t}{\sqrt{1+\left( \frac{g\,t}{c} \right)^2}} .
\end{align}
\end{subequations}
The Lorentz factor also grows with time, according to
\begin{equation}
 \gamma(t) = \sqrt{1+\left( \frac{g\,t}{c} \right)^2} .
\end{equation}
At very late time $t\gg c/g$ the trajectory can be approximated by
\begin{subequations}
\begin{align}
 x & \approx sign(g) \, c\,t \\
 v & \approx sign(g) \, c \\
 \gamma & \approx \frac{|g|\,t}{c} .
\end{align}
\end{subequations}
The particle quickly moves at ultra-relativistic speeds with a Lorentz factor growing almost linearly with time. Motions along the $y$ and $z$ directions are easily found with similar expressions. For an arbitrary direction of the electric field, if its strength remains constant, the final Lorentz factor remains the same for all simulations if the final time measured in units of $c/g$ remains identical.

As a diagnostic of the performances of the algorithm, we compute the relative error in the final value of the distance left from the origin and the final Lorentz factor for different time steps $\Delta t$ and different electric field strengths. For concreteness, the final time is fixed to $10^9\,c/g$ such that the final Lorentz factor is close to $\gamma_f=10^9$. The time step is also measured in units of $c/g$. The final distance from the origin is estimated from the trajectory and equal to $\{10^{12}, 10^9, 10^6\}$ for respectively intensities of $E=\{10^{-3},1,10^3\}$. A summary of final errors for an electric field equally directed along all three axes is given in table~\ref{tab:Electric}. The relative error is insensitive to the normalized electric field strength. It is computed for any quantity~$X$ according to
\begin{equation}
 \epsilon_X = \frac{|X_{\rm sim} - X_{\rm th}|}{|X_{\rm th}|}
\end{equation}
where $X_{\rm sim}$ is the value extracted from the simulations and $X_{\rm th}$ the theoretical value. In the table~\ref{tab:Electric} we only show the results for an electric field strength $E=1$.
\begin{table}
  \begin{center}
\def~{\hphantom{0}}
  \begin{tabular}{ccc}
  \hline
      $\log(\Delta t)$  & $\epsilon_d$ & $\epsilon_\gamma$ \\
  \hline
       ~3 & 4.989904e-07 & 1.065192e-11 \\
       ~2 & 4.910838e-08 & 1.001552e-10 \\
       ~1 & 3.184323e-09 & 8.975194e-10 \\
       ~0 & 1.318456e-08 & 1.309557e-08 \\
       -1 & 5.514039e-08 & 5.514122e-08 \\
       -2 & 5.556778e-07 & 5.556774e-07 \\
       -3 & 1.489058e-05 & 1.489036e-05 \\
  \hline
  \end{tabular}
  \caption{Relative precision for the final position of the particle and for the Lorentz factor in an uniformly accelerating normalized electric field directed along the axis $\ex+\ey+\ez$. The particle was initially at rest and its final Lorentz factor is $\gamma_f=10^9$.}
  \label{tab:Electric}
  \end{center}
\end{table}
For any value of the time step even when $\Delta t \gg c/g$ the final Lorentz factor is computed to high accuracy with 11 digits for the largest time step. Decreasing $\Delta t$ will increase the number of operations to be done and therefore lower the precision. However the final position is less sensitive to $\Delta t$ given with 7-8 digits of precision except for $\Delta t = 10^{-3} \, c/g$ where only 5 digits of precision are computed. Obviously, reducing the time step increase the accuracy of the position but the number of iterations also increases and the relative error in the Lorentz factor increases too. For $\Delta t=10^{-3}$ the distance is given with 5~digits and the Lorentz factor also thus 3 digits less for distance and 6 digits less for velocity. Our algorithm is able to catch particles with very high Lorentz factors even with large time steps. A compromise between precision in the position and velocity space and computational cost requires an appropriate choice of the time step.

\begin{figure}
\centering
\input{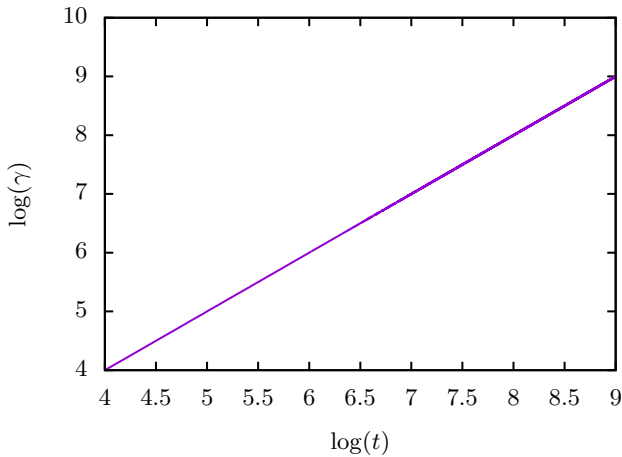}
\caption{Acceleration of an electron in a constant electric field along the $x$ axis with $E=10^3$. The Lorentz factor is plotted versus time in a log-log scale.}
\label{fig:acceleration_E_x}
\end{figure}

To conclude on pure electric acceleration, an example is provided in fig.\ref{fig:acceleration_E_x} where the Lorentz factor is shown to increase linearly with time. There are no bounds for the particle energy and the code will at some stage fail to reproduce Lorentz factor above $10^{15}$ corresponding to double precision representation of floating numbers.

\subsection{Constant magnetic field}

The second instructive example assumes a constant magnetic field along the $z$ axis. The relativistic trajectory of a charged particle is well known and given by an helicoidal motion corresponding to a constant speed~$v_z$ in the $z$ direction and a circular path in the plane perpendicular to the magnetic field lines. Therefore the motion is described by
\begin{subequations}
\begin{align}
 x & = r_B \, \cos \omega_B \, t \\
 y & = r_B \, \sin \omega_B \, t \\
 z & = v_z \, t
\end{align}
\end{subequations}
where the Larmor radius of the circular part is $r_B=v_\perp/\omega_B$, the synchrotron frequency is $\omega_B=q\,B/\gamma\,m$ and appropriate initial conditions for the position and velocit should be used. The Lorentz factor is obviously constant because the magnetic field does not work and equal to
\begin{equation}
\label{eq:Gamma}
 \gamma = \left( 1 - \frac{v_\perp^2+v_z^2}{c^2} \right)^{-1/2} .
\end{equation}
The solutions for a magnetic field directed along the $x$ and $y$ axis are obtained by a permutation of the components of the position vector. The most general configuration requires magnetic field components along all three axis.

\begin{figure}
\centering
\input{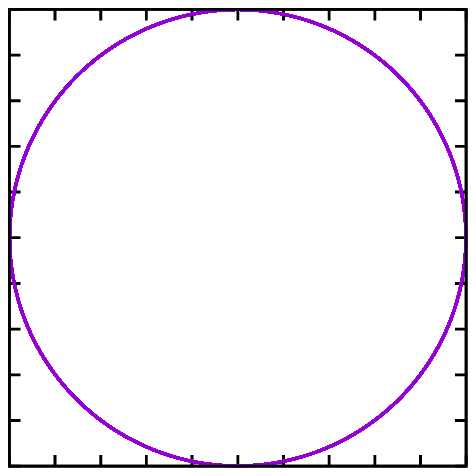}
\caption{Gyromotion of an electron in a constant magnetic field along the $z$ axis with $v_\perp=0.999\,c$ and $\gamma=10^6$.}
\label{fig:Gyromotion_B_z}
\end{figure}

\begin{figure}
\centering
\input{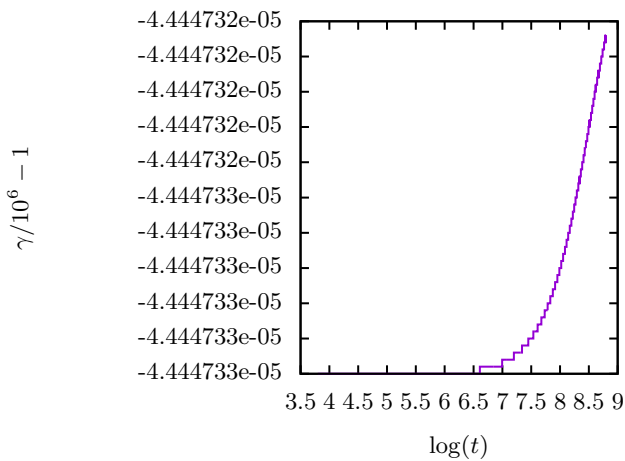}
\caption{Lorentz factor of an electron in a constant magnetic field along the $z$ axis with $v_\perp=0.999\,c$ and $\gamma=10^6$. The Lorentz factor is plotted versus time in a log-log scale and remains constant.}
\label{fig:Gyromotion_B_z_gamma}
\end{figure}

For the numerical tests, the time step~$\Delta t$ is given in units of $2\pi/\omega_B$ and the particle makes 100~turns. For the numerical parameters, we used the same field intensities and time steps as for the constant electric field case. The perpendicular speed is $v_\perp=0.999\,c$ and the constant Lorentz factor is set to $\gamma=10^6$ from which we deduce $v_z$ according to eq.~(\ref{eq:Gamma}). An example of gyromotion is shown in fig.~\ref{fig:Gyromotion_B_z}. The trajectory remains perfectly circular, there is no loss of energy and no shrinking of the orbit after 100~turns. The Lorentz factor is shown in fig.~\ref{fig:Gyromotion_B_z_gamma} and is perfectly constant during the full time of the run and equal to its initial value of $10^6$ to very good accuracy.
As a more precise check of the algorithm, we compute the relative errors in the distance travelled along the magnetic axis and the final Lorentz factor as shown in table~\ref{tab:Magnetic}. In this run, we assume a constant magnetic field along the vector $\ex+\ey+\ez$.
\begin{table}
  \begin{center}
\def~{\hphantom{0}}
  \begin{tabular}{ccc}
  \hline
      $\log(\Delta t)$  & $\epsilon_d$   &   $\epsilon_\gamma$ \\
  \hline
       ~3 & 1.005862e-13 & 3.095302e-13 \\
       ~2 & 1.005862e-13 & 3.095302e-13 \\
       ~1 & 1.005862e-13 & 3.095302e-13 \\
       ~0 & 1.005862e-13 & 3.095302e-13 \\
       -1 & 1.005862e-13 & 3.095302e-13 \\
       -2 & 1.005862e-13 & 3.095302e-13 \\
       -3 & 1.005862e-13 & 3.095302e-13 \\
  \hline
  \end{tabular}
  \caption{Relative precision for the final position of the guiding centre and for the Lorentz factor in a constant magnetic field directed along the axis $\ex+\ey+\ez$. The Lorentz factor is initially $\gamma=10^6$.}
  \label{tab:Magnetic}
  \end{center}
\end{table}
Whatever the time step, the position of the guiding centre is given with almost full accuracy of double precision, 13~digits. The Lorentz factor also remains constant within 13~digits of precision. This is achieved for any time step. Longer integration time does not degrade this accuracy.

\subsection{Cross electric and magnetic fields}

\begin{figure}
\centering
\input{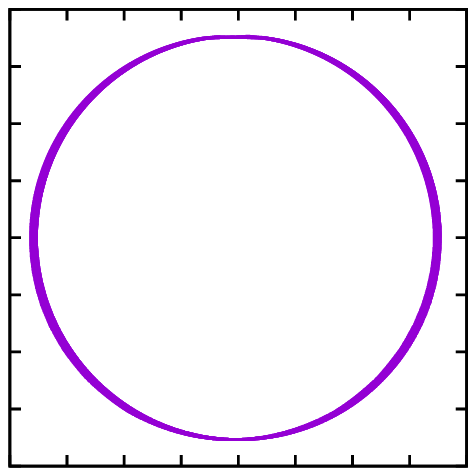}
\caption{Gyromotion of an electron in the electric drift frame with $\Gamma_{\rm E} \approx 707.107$ and $\gamma=100$ in the electric drift frame. 10~rotations are plotted.}
\label{fig:Gyromotion_Drift_E}
\end{figure}

\begin{figure}
\centering
\input{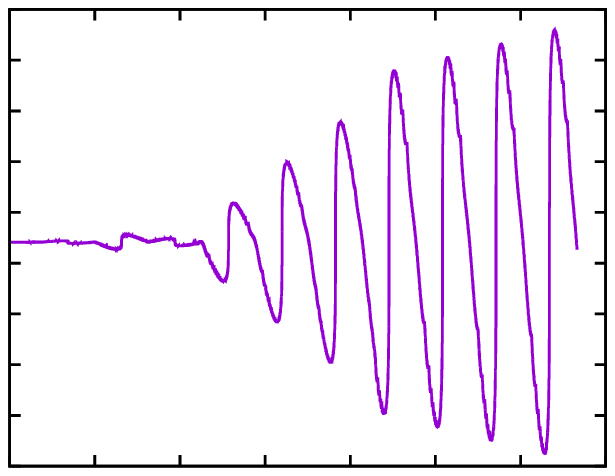}
\caption{Radius of the orbit of an electron in the electric drift frame with $\Gamma_{\rm E} \approx 707.107$.}
\label{fig:Gyromotion_Drift_E_rB}
\end{figure}

\begin{figure}
\centering
\input{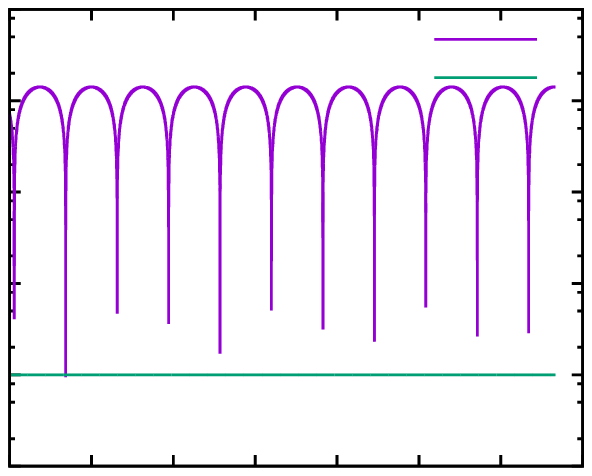}
\caption{Lorentz factor of an electron in the observer frame and in the drift frame with $\Gamma_{\rm E} \approx 707.107$.}
\label{fig:Gyromotion_Drift_E_gamma}
\end{figure}

As another proof of the efficiency of the algorithm, we compute the trajectories in a crossed electromagnetic field ($\mathbf{E} \cdot \mathbf{B} = 0$) where the average motion is an electric drift in the $\mathbf{E} \wedge \mathbf{B}$ direction at the electric drift speed $\mathbf{v}_{\rm E} = \mathbf{E} \wedge \mathbf{B}/B^2$. In the frame moving at $\mathbf{v}_{\rm E}$ the electric field vanishes and the particle simply follows an helicoidal motion along the constant magnetic field, see the case treated in the previous paragraph. This motion is only allowed for weak electric fields satisfying $E<c\,B$ thus enforcing $v_{\rm E}<c$. For concreteness, let us assume an electric field directed along $\ey$ and a magnetic field directed along $\ez$. The electric drift speed becomes $\mathbf{v}_{\rm E} = (E_y/B_z) \, \ex$. A Lorentz transformation of the electromagnetic field with Lorentz factor $\Gamma_{\rm E} = 1/\sqrt{1-v_{\rm E}^2/c^2}$ along $\mathbf{v}_{\rm E}$ shows that in the comoving frame
\begin{subequations}
\begin{align}
 \mathbf{E}' & = 0 \\
 \mathbf{B}' & = \mathbf{B}/\Gamma_{\rm E} .
\end{align}
\end{subequations}
As initial conditions for the particle position and velocity, we choose a helicoidal motion as explained in the previous paragraph and corresponding to an evolution in the magnetic field $\mathbf{B}'$ as seen in the electric drift frame. These quantities are then transformed to the observer inertial frame according to Lorentz transformations for the three velocity $\mathbf{v}$ of the particle. The algorithm is checked by computing the particle trajectory in the drift frame and the corresponding Lorentz factor of the particle that should remain constant in that frame. Using the Lorentz transformation the coordinates in the drift frame are
\begin{subequations}
\begin{align}
 x' & = \Gamma_{\rm E} \, ( x - v_{\rm E} \, t ) \\
 y' & = y \\
 z' & = z .
\end{align}
\end{subequations}
For numerical purposes, the intensity of the electric field is set to $E=0.999999$ and that of the magnetic field is $B=1$ leading to a Lorentz factor of the electric drift motion of $\Gamma_{\rm E} \approx 707.107$. The Lorentz factor in the electric drift frame is fixed to $\gamma=100$. Typical results are depicted in fig.~\ref{fig:Gyromotion_Drift_E} for the trajectory in the electric drift frame which is usually an helicoidal motion and here exactly a circle in the comoving plane $x'0'y'$. The trajectory projected onto the $x'O'y'$ plane remains a circle to very good accuracy with a relative change in radius less that 1\%. Fig.~\ref{fig:Gyromotion_Drift_E_rB} supports this fact. The Lorentz factor is shown in fig.~\ref{fig:Gyromotion_Drift_E_gamma}. In the observer frame it reaches values up to $\gamma\approx10^5$ whereas in the electric drift frame it is computed according to $\gamma=\Gamma_E \, \Gamma \, (1 - \boldsymbol\beta_E \cdot \boldsymbol\beta)$. As a check of our algorithm, we plot these Lorentz factors on the same fig.~\ref{fig:Gyromotion_Drift_E_gamma} and indeed retrieve $\gamma=100$ within 8~digits of precision.

\begin{figure}
\centering
\input{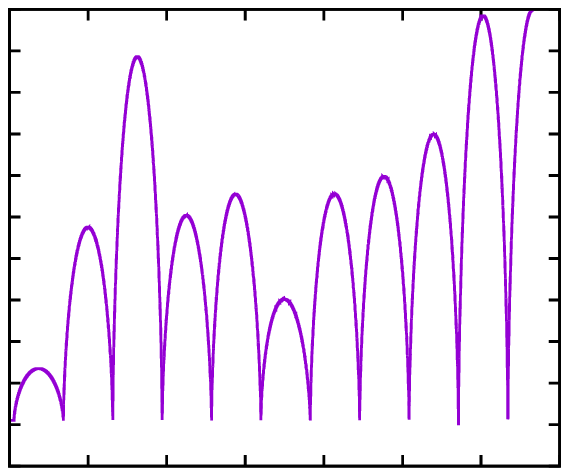}
\caption{Radius of the orbit of an electron in the electric drift frame with $\Gamma_{\rm E} \approx 70.71$.}
\label{fig:Gyromotion_Drift_E_rB_2}
\end{figure}

The radius of the circular motion suffers from a slight oscillation amplitude increase with time. The electric drift regime in the relativistic limit represents a severe test for any relativistic particle pusher. Indeed, for lower electric drift speed $v_{\rm E}$, the circular orbit is better conserved like for instance taking $\Gamma_{\rm E} \approx 70.71$ (E=0.9999), see fig.~\ref{fig:Gyromotion_Drift_E_rB_2}.

The most general and realistic fields are spatially and temporally varying. In these cases, the electromagnetic field has to be determined at some position and time during the particle motion. As severe tests of our algorithm, we study four problems of which three have exact analytical relativistic solutions for the particle trajectory. The relativistic electrostatic Keplerian two body problem of a electron orbiting around a fixed positive ion is an interesting test for a spatially varying electric field. It is a central force case. The other two solutions correspond to a particle moving in a plane electromagnetic wave linearly or circularly polarized. A last less trivial example is depicted by a particle drifting in the equatorial plane of a static magnetic dipole. We discuss in depth these regimes in the following paragraphs. The analytical solutions are described in \cite{uzan_theories_2014} and for completness we recall them in the following paragraphs.

\subsection{Central electric force}

\begin{figure}
\centering
\input{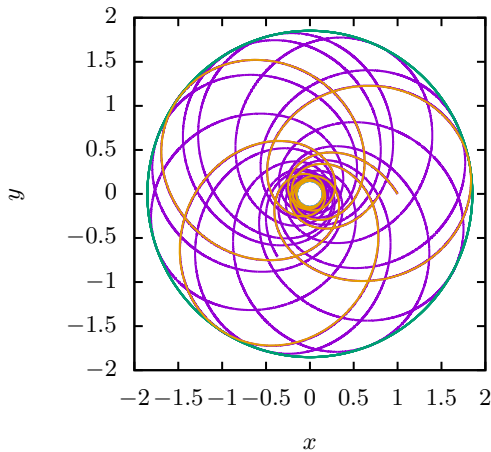}
\caption{Motion of an electron in the electric field of a fixed proton. The relativistic precession of the orbit is clearly visible. The minimal and maximum radius of the orbit as predicted by the analytical formulae are shown by two circles tangent to the trajectory. A piece of the exact analytical solution is also shown in yellow.}
\label{fig:E_kepler}
\end{figure}

\begin{figure}
\centering
\input{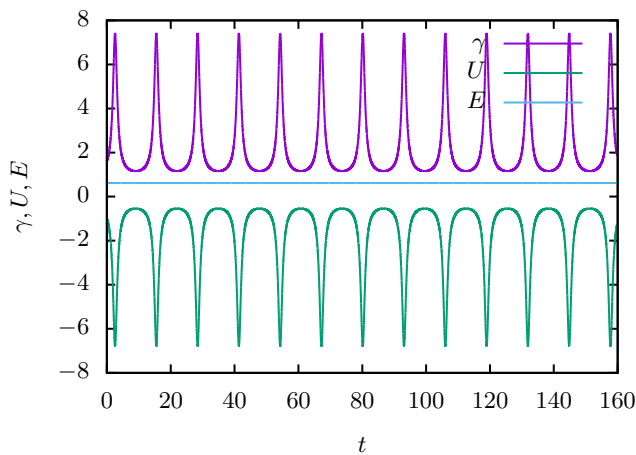}
\caption{Total energy~$E$, relativistic kinetic energy~$\gamma\,m\,c^2$ and electrostatic potential energy~$U$ of an electron in the electric field of a fixed proton. $E$ is perfectly conserved, being a constant of motion.}
\label{fig:E_kepler_energie}
\end{figure}

The two body problem in gravitational physics can be transposed in an equivalent electrostatic problem including relativistic corrections to the velocity. In this case, a particle with charge $q$ and mass $m$ orbits around a fixed central particle with charge~$Q$. For bounded orbits we require $q\,Q<0$. Solutions are given by conservation of energy~$E$ and angular momentum~$L$. The electric force applied to the orbiting particle is 
\begin{equation}
 f = \frac{q\,Q}{4\,\pi\,\varepsilon_0\,r^3} \, \mathbf{r} .
\end{equation}
The orbital motion stays in a plane that we choose as the $xOy$ plane. In polar coordinates $(r,\phi)$ the conservation laws for angular momentum and energy are translated into
\begin{subequations}
\begin{align}
 \gamma \, \dot\phi & = \frac{L}{m\,r^2} \\
 \gamma \, m \, c^2 & = E - \frac{q\,Q}{4\,\pi\,\varepsilon_0\,r} .
\end{align}
\end{subequations}
Eliminating the Lorentz factor $\gamma$ from these expressions, the equations of motion in polar coordinates become 
\begin{subequations}
\begin{align}
 \dot \phi & = \frac{L\,c^2}{r^2 \, \left( E - \frac{q\,Q}{4\,\pi\,\varepsilon_0\,r} \right)} \\
 \dot r^2 & = c^2 \, \left[ 1 - \frac{L^2\,c^2/r^2 + m^2\,c^4}{\left( E - \frac{q\,Q}{4\,\pi\,\varepsilon_0\,r} \right)^2} \right] .
\end{align}
\end{subequations}
For a circular orbit $\dot r=0$ and $\dot \phi=\Omega$. Equality between centrifugal force and electrostatic attraction induces an angular rotation rate satisfying
\begin{equation}
 \Omega^4 + \left( \frac{q\,Q}{4\,\pi\,\varepsilon_0\,m\,r} \right)^2 \, \frac{\Omega^2}{r^2\,c^2} - \left( \frac{q\,Q}{4\,\pi\,\varepsilon_0\,m\,r} \right)^2 \, \frac{1}{r^4} = 0 .
\end{equation}
The solution is 
\begin{equation}
\Omega = - \frac{q\,Q}{4\,\pi\,\varepsilon_0\,m\,r^2\,c\,\sqrt{2}} \, \left[ -1 + \sqrt{1+ \frac{4\,m^2\,c^4}{\left( \frac{q\,Q}{4\,\pi\,\varepsilon_0\,r} \right)^2}} \right]^{1/2} .
\end{equation}
In the general case of an elliptic bound orbit, dividing $\dot r^2$ by $\dot \phi^2$ and setting $u=1/r$ we find a ordinary differential equation
\begin{equation}
 \frac{d^2u}{d\varphi^2} + \left[ 1 - \left( \frac{q\,Q}{4\,\pi\,\varepsilon_0\,L\,c} \right)^2 \right] u = - \frac{q\,Q}{4\,\pi\,\varepsilon_0\,c^2} \, \frac{E}{L^2} .
\end{equation}
For $\left|\frac{q\,Q}{4\,\pi\,\varepsilon_0\,L\,c}\right|<1$, the general solution is 
\begin{subequations}
\begin{align}
 r & = \frac{p}{1 + e \, \cos(\Omega_p \, (\varphi - \omega))} \\
 \Omega_p & = \sqrt{1 - \left( \frac{q\,Q}{4\,\pi\,\varepsilon_0\,L\,c}\right)^2 } \\
 p & = \frac{\Omega_p^2}{- \frac{q\,Q}{4\,\pi\,\varepsilon_0\,c^2} \, \frac{E}{L^2}} \\
 e^2 & = \frac{1}{E^2} \, \left[ m^2\,c^4 + \frac{E^2 - m^2\,c^4}{\left( \frac{q\,Q}{4\,\pi\,\varepsilon_0\,L\,c}\right)^2} \right] .
\end{align}
\end{subequations}
An example of relativistic particle trajectory showing the precession of the orbit is given in fig.~\ref{fig:E_kepler}. A piece of the exact analytical solution is also shown and matches perfectly the output of the numerical simulations. The total energy~$E$ is split into relativistic kinetic energy~$\gamma\,m\,c^2$ and electrostatic potential energy~$U$. Inspection of fig.~\ref{fig:E_kepler_energie} demonstrates that the total energy is exactly conserved during time evolution.

\subsection{Magnetic dipole}

\begin{figure}
\centering
\input{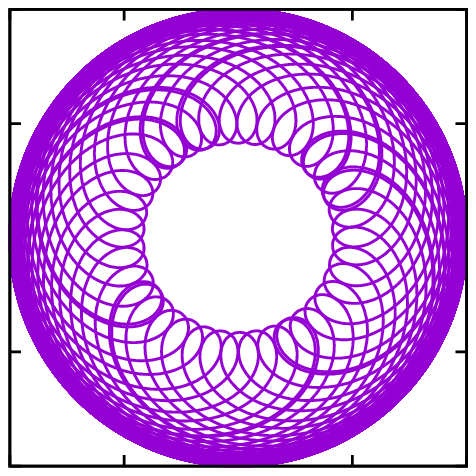}
\caption{Relativistic drift motion of an electron in a magnetic dipolar field.}
\label{fig:Drift_B_dipole}
\end{figure}

\begin{figure}
\centering
\input{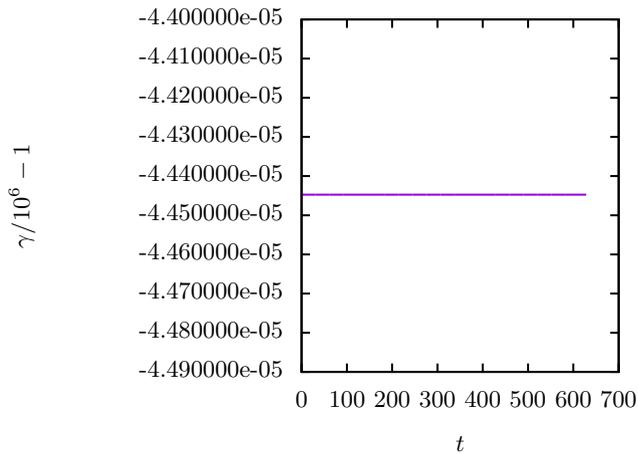}
\caption{Lorentz factor of the drift motion of an electron in a magnetic dipolar field.}
\label{fig:Drift_B_dipole_gamma}
\end{figure}

In the equatorial plane of a magnetic dipole, the particle gyrates around the origin where the dipole is located. This motion is induced by the magnetic gradient drift in the azimuthal direction. The particle stay within two circles. An example of magnetic drift motion is shown in fig.~\ref{fig:Drift_B_dipole} for $\gamma=10^6$. The Lorentz factor is conserved as checked in fig.~\ref{fig:Drift_B_dipole_gamma}.

We finish our extensive test of the algorithm by considering two time varying electromagnetic fields represented by linearly and circularly polarized plane waves.

\subsection{Linearly polarized plane wave}

\begin{figure}
\centering
\input{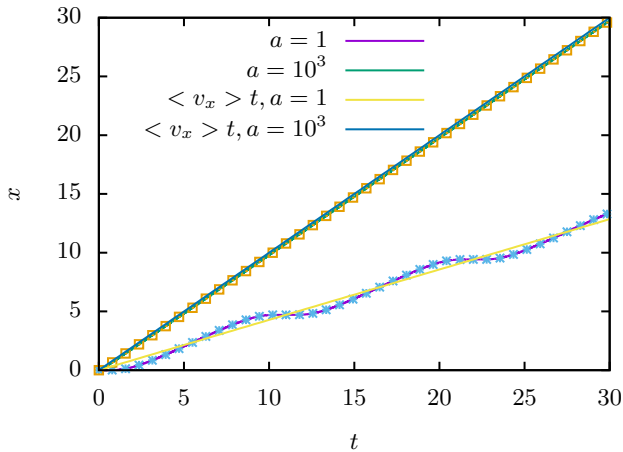}
\caption{Motion of an electron in a linearly polarized plane wave for different strength parameters $a=1,10^3$. Symbols correspond to the analytical solution eq.~(\ref{eq:position_onde_lineaire}).}
\label{fig:onde_lineaire}
\end{figure}

\begin{figure}
\centering
\input{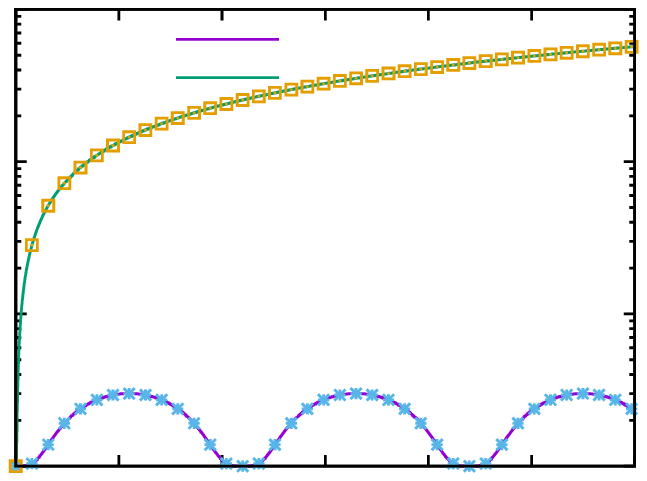}
\caption{Lorentz factor of an electron in a linearly polarized plane wave for different strength parameters $a=1,10^3$. Symbols correspond to the analytical solution eq.~(\ref{eq:vitesse_onde_lineaire}).}
\label{fig:onde_lineaire_gamma}
\end{figure}

Consider a linearly polarized plane wave propagating along the $\ex$ direction such that the vector potential is $A^\alpha=(0,0,\frac{E}{\omega}\,\cos\xi,0)$. The wave vector is therefore $(\frac{\omega}{c},k,0,0)$ from which we deduce the phase $\xi=\omega\,t-k\,x$. The electromagnetic field is then given by 
\begin{subequations}
\begin{align}
\mathbf{E} & = E \, \sin \xi \, \ey \\
\mathbf{B} & = \frac{E}{c} \, \sin \xi \, \ez .
\end{align}
\end{subequations}
Initially the particle is at rest with a 4-velocity $u^\alpha_0=(c,\mathbf 0)$. Introducing the strength parameter of the wave by the ratio
\begin{equation}
a = \frac{q\,E}{m\,c\,\omega}
\end{equation}
the 4-velocity will have components
\begin{subequations}
\label{eq:vitesse_onde_lineaire}
\begin{align}
 u^x & = \frac{a^2}{2} \, c \, (\cos \xi - 1 )^2 \\
 u^y & = -a \, c \, (\cos \xi - 1 ) \\
 u^0 & = c + u^x .
\end{align}
\end{subequations}
The mean spatial velocity becomes 
\begin{subequations}
\begin{align}
 <v^x> & = \frac{3}{4} \, \frac{a^2\,c}{1+3\,a^2/4} \\
 <v^y> & = \frac{a\,c}{1+3\,a^2/4} .
\end{align}
\end{subequations}
After integration, assuming the particle starts at rest at the origin at $\xi=0$, we find
\begin{subequations}
\label{eq:position_onde_lineaire}
\begin{align}
 \omega \, x & = \frac{a^2\,c}{8} \, (6\,\xi - 8 \, \sin\xi + \sin 2\,\xi) \\
 \omega \, y & = a\,c\, (\xi - \sin\xi) \\
 \omega \, c \, t & = c\,\xi + \omega \, x .
\end{align}
\end{subequations}
Examples of motion along the $x$ axis are shown in fig.~\ref{fig:onde_lineaire} for a mildly $a=1$ and an ultra-relativistic $a=10^3$ strength parameter. The mean motion with average velocity $<v_x>$ is also shown. The associated Lorentz factor evolution is shown in fig.~\ref{fig:onde_lineaire_gamma}. The numerical integration is compared to the analytical solution depicted by coloured symbols.

\subsection{Circularly polarized plane wave}

\begin{figure}
\centering
\input{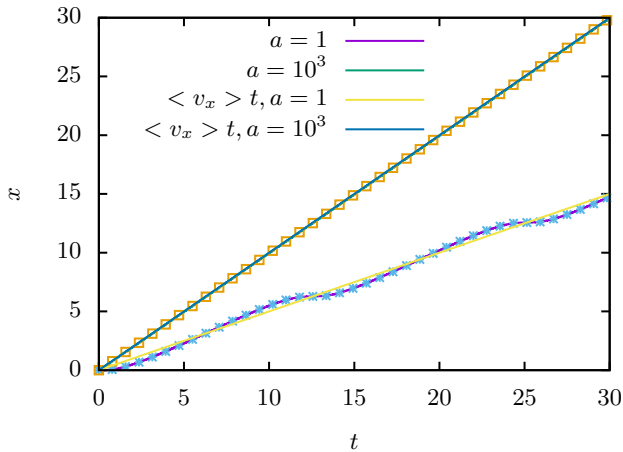}
\caption{Motion of an electron in a circularly polarized plane wave for different strength parameters $a=1,10^3$. Symbols correspond to the analytical solution eq.~(\ref{eq:position_onde_circulaire}).}
\label{fig:onde_circulaire}
\end{figure}

\begin{figure}
\centering
\input{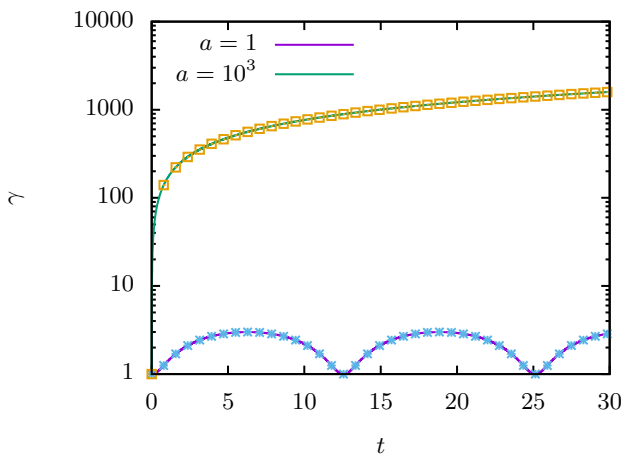}
\caption{Lorentz factor of an electron in a circularly polarized plane wave for different strength parameters $a=1,10^3$. Symbols correspond to the analytical solution eq.~(\ref{eq:vitesse_onde_circulaire}).}
\label{fig:onde_circulaire_gamma}
\end{figure}

Consider a circularly  polarized plane wave propagating in the $\ex$ direction such that the vector potential has components $A^\alpha=(0,0,\frac{E}{\omega}\,\cos\xi,\frac{E}{\omega}\,\sin\xi)$ and the wave vector $(\frac{\omega}{c},k,0,0)$ thus the phase $\xi=\omega\,t-k\,x$. The electromagnetic field is then given by 
\begin{subequations}
\begin{align}
\mathbf{E} & = E \, ( \sin \xi \, \ey - \cos \xi \, \ez ) \\
\mathbf{B} & = \frac{E}{c} \, ( \sin \xi \, \ez + \cos \xi \, \ey ) .
\end{align}
\end{subequations}
Initially the particle is at rest with 4-velocity $u^\alpha_0=(c,\mathbf 0)$. The time evolution of the components of this 4-velocity will be
\begin{subequations}
\label{eq:vitesse_onde_circulaire}
\begin{align}
 u^x & = a^2 \, c \, (1 - \cos \xi ) = a \, u^y \\
 u^y & = a \, c \, (1 - \cos \xi ) \\
 u^z & = - a \, c \, \sin \xi \\
 u^0 & = c + u^x .
\end{align}
\end{subequations}
The mean spatial velocity becomes 
\begin{subequations}
\begin{align}
 <v^x> & = \frac{a^2\,c}{1+a^2} \\
 <v^y> & = \frac{a\,c}{1+a^2} \\
 <v^z> & = 0 .
\end{align}
\end{subequations}
After integration, assuming the particle starts at rest at the origin at phase $\xi=0$, we find
\begin{subequations}
\label{eq:position_onde_circulaire}
\begin{align}
 \omega \, x & = a^2\,c \, (\xi - \sin\xi) \\
 \omega \, y & = a\,c\, (\xi - \sin\xi) \\
 \omega \, z & = a\,c\, (\cos\xi-1) \\
 \omega \, c \, t & = c\,\xi + \omega \, x .
\end{align}
\end{subequations}
Examples of motion along the $x$ axis are shown in fig.~\ref{fig:onde_circulaire} for a mildly $a=1$ and an ultra-relativistic $a=10^3$ strength parameter. The mean motion with average velocity $<v_x>$ is also shown. The corresponding evolution of the Lorentz factor is given in fig.~\ref{fig:onde_circulaire_gamma}. The numerical integration is compared to the analytical solution depicted by coloured symbols.

\section{Conclusion}

We developed an efficient and fully implicit algorithm to solve the Lorentz force equation for ultra-relativistic particles reaching Lorentz factors higher than $10^{9}$. Although it requires matrix multiplication, this additional computational cost is completely compensated by the implicit nature of the scheme and its very good behaviour and accuracy for ultra-relativistic motion on large time scales for spatially and temporally varying fields. These features are compulsory for any study of plasmas in high-energy astrophysics where such extreme conditions are easily met like for instance in neutron stars and especially pulsars for which the ratio between the Larmor frequency and the neutron star spin frequency is huge, spanning more than 10 decades. Several more important tests have been envisaged like motion in a Coulombian electric field, a kind of relativistic Kepler two-body problem thus in an inhomogeneous electric field. Motion in a linearly or circularly polarized plane wave represents also a challenging test of the algorithm in extreme intense electromagnetic fields. Applications are not restricted to high-energy astrophysics, interaction of high-intensity laser fields with plasmas is another fruitful domain to investigate.

Our implicit algorithm updates the velocity from an evaluation of the electromagnetic field at some point in spacetime, chosen as $t^*,\mathbf{r}_*$, eq.~(\ref{eq:Picard}). Other schemes could be tested such as an interpolation of the Lorentz force in such a way that
\begin{subequations}
\label{eq:Picard2}
\begin{align}
 \frac{\mathbf{r}^{n+1} - \mathbf{r}^n}{\Delta t} & =  \frac{\mathbf{v}^{n+1} + \mathbf{v}^n}{2} = \frac{1}{2} \, \left( \frac{\mathbf{u}^{n+1}}{\gamma^{n+1}} + \frac{\mathbf{u}^{n}}{\gamma^{n}}\right) \\
 \frac{\mathbf{u}^{n+1} - \mathbf{u}^n}{\Delta t} & = \frac{q}{2\,m} \, [ \mathbf{E}(\mathbf{r}^{n+1},t^{n+1}) + \mathbf{v}^{n+1} \wedge \mathbf{B}(\mathbf{r}^{n+1},t^{n+1}) + \mathbf{E}(\mathbf{r}^n,t^n) + \mathbf{v}^n \wedge \mathbf{B}(\mathbf{r}^n,t^n) ] .
\end{align}
\end{subequations}
Here again, the non-linear system is solved iteratively by Picard method. However, we leave this possibility for future work.

Following the discussion presented in the work by \cite{qiang_high_2017}, it is possible to use higher order integration schemes to reduce the computational costs to reach a prescribed accuracy. \cite{qiang_high_2017} technique can be applied to any time reversible second order relativistic integrator algorithm.

Ultra-relativistic particles in ultra strong magnetic fields are prone to damping via radiation reaction. This viscous force needs to be implemented in the equation of motion to catch the full dynamics of relativistic plasmas subject to intense radiation fields. This additional frictional force can be implemented straightforwardly with the fully implicit technique exposed in this paper but is left for future work.

\section*{Acknowledgements}

I am grateful to the referees for their constructive comments and suggestions. This work has been supported by the French National Research Agency (ANR) through the grant No. ANR-13-JS05-0003-01 (project EMPERE).


\end{document}